# Survey of Sybil Attacks in Social Networks

Rupesh Gunturu

*Abstract*

This paper reviews the Sybil attack in social networks, which has the ability to compromise the whole distributed network. In the Sybil attack, the malicious user claims multiple identities to compromise the whole network. Sybil attacks can be used to change the overall ranking in voting applications, bad-mouth an opinion, access resources or to break the trust mechanism behind a P2P network. In this paper, different defense mechanisms used to mitigate Sybil attacks are also reviewed.

**Introduction**

The concept of social networks, first investigated by sociologists in the 1960s, is now one of the key ways social computing is taking place. Social networking sites allow users to create personal profiles, share images and to connect with a large network of friends and often with a lot of strangers. Because of the wide variety of features such as games, puzzles, real time applications, image sharing, instant messaging, and so on, the personal information shared amongst a group can be compromised and this activity also leads to a loss of productivity. There are additionally attacks on the reputation framework associated with social networks with an eye on influencing the integrity of the data or the communication processes of the network [1]-[19].

The need for connections is partly driven by the uncertainty that underlies information structures [20],[21] within society. The network networked society [22],[23] must therefore find new ways to connect, which also leads it open to a variety of attacks [24],[44]. It also opens questions related to identity and anonymity. Some of these questions have been traditionally addressed by assigning random sequences to different individuals [25]-[30], as well as the use of cryptography based on quantum mechanics [31]-[34].

The main reason for threats in social networking is due to its open architecture and its susceptibility to different viruses [10]. Due to this open structure, the usual culprits such as spam, cross-scripting, social engineering can cause much damage. Thus an adversary may masquerade as an honest user and divert the honest user to other pages to spread the malware which leads to a phishing attack [18][39][41] or a malicious user may create multiple bogus identities to gain sensitive information from users [39][8]. This latter attack where an adversary creates multiple bogus identities to gain sensitive information like full name, SSN, bank details from users to do variety of cyber-crimes is the Sybil attack [43]. Sybil attacks are considered to be very crucial in distributed decentralized systems like social networks because a lot of personal information is shared openly and it is impossible to erase the information once uploaded [40].

In P2P networks, which go back to Napster [43], each task or problem is accomplished by subdividing the task among multiple nodes or peers. Unlike the client server-model, in which the central server provides the resources to the client requests (services provided by web servers and browsers act as clients), all the nodes in a P2P network act as sources and receivers [38]. The peers have their own share of resources such as bandwidth, power, download speed, routing channels to transfer data, computational resources etc. Each peer can communicate and also provide its resources to an external system without the need for central authority. A P2P network is built on the assumption that each entity in the network holds a single identity. When an adversary introduces many bogus identities with a single entity or with no entity at all, a Sybil attack occurs. Using Sybil identities, an adversary may provide false opinions for his/her evil benefits, limit the amount of resources reaching each node, break the trust mechanism in a P2Pnetwork and may even cause a Denial-of-Service attack (DoS).

In the initial researches to deal with Sybil attacks, network architectures were re-designed and secure mechanisms such as digital signatures and digital analyzers were used to mitigate the Sybil attacks. Much effort has gone into the study of trust relationships in social networks [1][2][45][47] and community based schemes to reduce the influences of Sybil attacks [6][47]. This paper reviews Sybil detection schemes based on behavior attributes of Sybil users. It also discusses classification of Sybil attacks, examples of Sybil attacks, social network based Sybil defenses, behavior based Sybil defenses followed by conclusion & references.

## Classification of Sybil attacks

### (i) Direct vs. In-Direct communication
To launch a Sybil attack in a distributed network, the attacker must consider the type of communication between honest nodes and Sybil nodes [5],[42]. If the communication between honest node and Sybil node is direct, i.e. if the attacker can directly communicate with the honest node using fake identities, it is a case of *direct communication*. However, if the attacker has to use his legitimate identity to communicate with the honest node, and then divert the Sybil data to the honest node via the legitimate node, it is the case of *indirect communication*. It is easier for the attackers to launch Sybil attacks in case of direct communication and it is also more difficult to detect such attacks.

### (ii) Busy vs. Idle
In a P2P network, normally, only few Sybil identities participate in the network while the others remain idle. The power of the Sybil attacker comes from the number of identities he or she holds. If an attacker could afford to get fake identities easily, he or she can make the identities appear more realistic by making them leave and join the network multiple times pretending as an honest node. However, if the number of the Sybil identities are limited, the Sybil identities must participate simultaneously to launch an attack.

### (iii) Simultaneous vs. Non Simultaneous
An adversary can create all Sybil nodes simultaneously or introduce them one by one. If the attacker introduces one node at a time and manages to establish different properties for different Sybil nodes, the chances of detecting the Sybil nodes in a P2Pnetwork becomes very difficult. However, the attacking time and complexity increases when nodes are introduced at different instances.

A simultaneous attack can be performed by involving all the Sybil identities simultaneously or a single physical node can change its identities in regular time slots to appear like all the identities are involved simultaneously.

In non-simultaneous attack, an attacker may bring all his identities into the network slowly over a period of time involving only few identities each time. This can be done by pretending that one identity is leaving the network while the other identity is joining the network. As honest identities generally tend to leave and join the network number of times, the malicious node won't be suspected if they pretend to leave or join the network now and then using different identities. The attacker can also use a number of physical devices to get different identities and can then switch among these identities to perform the attack.

**(iv) Insider vs. Outsider**
The impact of the Sybil attack depends on whether the attacker is inside or outside the distributed network. If the adversary is part of the network and holds at least one real identity, then the attacker is called an Insider, otherwise he or she is an outsider. An insider may introduce many fake identities, and pretend to communicate with other nodes using his fake identities. However, for an outsider, it is difficult to introduce Sybil identities into the network, as the distributed network system generally employs some kind of authentication procedure such as passwords, secret codes or encryption processes to access the system. An insider can transmit the false information over the network cloud or receive information from other nodes as the network generally trusts all its internal nodes. However, a Sybil node can easily be detected by monitoring the claimed communication between the suspect node and other nodes.

**Examples of Sybil attacks**

**(i) Mobile Ad-hoc Networks**
A Vehicular ad-hoc network (VANET) is a mobile network in which each moving car represents a node [9][15]. VANET treats each car as a router and allows cars within 100 to 300 meters range to connect with each other, creating a large mobile network. This technology is soon expected to be integrated with police and fire vehicles to connect with each other for safety [43]. In VANET, each node communicates with the nearest base station and with other mobile nodes for updates. Safety is considered to be the highest priority in mobile ad-hoc networks. Due to the weak authentication procedures in VANET, a malicious node may create multiple bogus identities and masquerade as an honest node. These fake identities may be used to misguide other nodes about a traffic jam in a particular routing path for her selfish desire to enjoy better traffic [13]. Secondly, the base station continuously relies on the votes of the nodes to generate a traffic status report. An adversary may use her Sybil vehicles to vote according to her needs, leading to a change in the actual traffic report. Thirdly, in VANET's safety warning messages are generated and sent from one node to another in case of traffic jams or during any unusual incidents to caution the vehicles to slow down. However, when the caution message reaches Sybil identities, the message may be dropped and other nodes may have to face danger. In addition, the Sybil vehicles may overload the base stations with too many requests and can cause a denial of service attack (DOS).

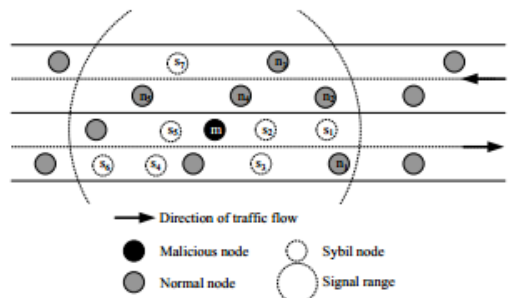

*Figure 1: Sybil attacks in VANET [9]*

**(ii) Distributed voting system**
In distributed decentralized systems like P2P networks, voting is used during various instances such as resource allocation, misbehavior detection etc [14]. The network believes that any entity in the network holds only a single identity and can vote only once. However, when an adversary creates bogus identities, she can vote multiple times and out-vote the honest user's decision according to her selfish needs [11]. As, the counting mechanism is indirect and it cannot verify whether the voting is done from a malicious node or an honest node, the actual decision of honest users is compromised.

In eBay, ratings to a seller are given according to the reviews written by the users. To increase the ratings of a particular merchant or to get a better rank in the Google page rank, Sybil attacks are used to write positive reviews about a particular merchant or product to mislead other users [43].

### (iii) Distributed Hash Tables (DHTs)

Distributed hash tables (DHT's) are used in large P2P networks to distribute data among individual nodes. In DHT, each node is given a unique ID or hash which acts a pointer to the node [14]. Each hash function consists of a (key, value) pair which could be used to access or retrieve the data from a node. In a P2P network with fixed number of nodes, a hash value falling in between two nodes is assigned to the nearest node. DHT's have the capability to look for any data item in the network and divide the responsibility of data storage among nodes. It can be used to form complex services such as Domain Name Services (DNS), Multicast, Instant messaging and also P2P file sharing.

Consider an Instant Messaging (IM) application in P2P networks which uses DHT routing [35], where a node wants to talk to other node by querying intermediate nodes. If an adversary creates fake identities and acts as a node in the P2P network, the queries from other nodes can be dropped or false information can be transmitted. If an adversary identifies the hash key of a particular node, he can overload the node by generating multiple hash keys consisting of (key, value) pairs. In addition, an adversary may create many Sybil IDs near a node ID and intercept the queries reaching a node.

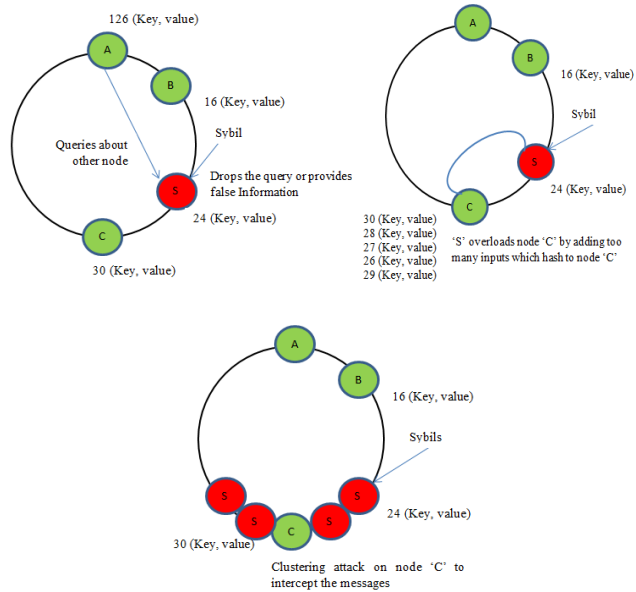

*Figure 2: Different attacks in DHT's*

### (iv) BitTorrent

BitTorrent, developed in the year 2001, is a protocol which is extremely popular for fast downloads of audio and video files [14]. It is a decentralized P2Pcontent distribution system which works on a mutual trading basis i.e. a user have to send files to other peers in order to receive files from them. It has been estimated that BitTorrent accounts for about 35% of the total traffic over the internet [17].

In BitTorrent protocol, a large file is broken into number of pieces, and each piece is identified by a unique hash function allocated to it [17]. In order to download a file, users have to download a ".torrent file" from websites such as private bay [14]. A torrent file consists of an URL of the tracker, an info section which tells us about the size of each piece, number of pieces, and the hash functions SHA1 associated with each piece [7],[17]. Tracker is a file which consists of a list of IP addresses which tells a user from which other peers he can download other pieces of the file. The protocol allows a user to download number of pieces simultaneously from different neighbors increasing the throughput. A user who is still downloading a file from the source is called a "Leecher" and a user who completed to download all the pieces of the file is called a "Seeder".

BitTorrent protocol works on a mutual trading basis i.e. a user has to transmit his pieces to other users in order to receive new pieces from them. Each BitTorrent client (a software program installed on the system) which can download and upload information, creating a P2P network, keeps track of the amount of data transmitted by the system in each stage and judges whether the machine can receive data or not in the consecutive stages [7]. By using multiple Sybil identities and pretending as multiple users, more pieces of the file could be downloaded than it is supposed to [34] [14]. In addition, the BitTorrent client installed on a computer can see the IP's downloading the same file. Using Sybil identities, an adversary may access the system to gain sensitive information like passwords, personal data or can even compromise the whole system drive [44].

**Countermeasures**
**(i) Resource Testing**
Douceur proposed resource testing [3] [42] for direct validation. In a P2P network, every node is believed to hold a single identity and is allocated its share of resources. The resource testing aims to test the resources of a node which include computation, storage or communication resources. In resource testing, a verifier assigns a certain task to the neighboring nodes and assigns them a threshold or time limit to respond. If the verifier listens responses in the stipulated time, they are considered to be honest, else called Sybil nodes.

Consider a wireless communication network, in which a node wants to detect the Sybil nodes around it. It is assumed that a radio is incapable of transmitting or receiving on multiple channels simultaneously [5]. A verifier who wants to detect the Sybil nodes around him assigns $n$ different channels to $n$ neighboring nodes and asks them to broadcast a message. If the verifier is able to listen to the messages in the channels he assigned, the node is considered to be honest otherwise it is considered as a Sybil node. The main disadvantage of resource testing is, each node wastes too much amount of its allocated resources to detect Sybils which is actually meant for a variety of productive purposes.

**(ii) Central trust authority**
Central trust party is one of the efficient Sybil defense schemes in which a third party is used to authenticate the nodes. In central trust authority, a node is said to be honest if it receives a certificate from the central server. The certificate may be a digital number [43] ,[44] or special hardware [5]. In P2P networks, a node has to be authenticated or certified by the central server before it can access the network or use its resources. As each and every step needs the central server involvement, it can be considered as one of the most reliable schemes to counter Sybil attacks. In addition, if the message transmitted has some kind of information or signature of the sender, the chances of Sybil attacks would be greatly reduced.

Though the central trust method seemed to be promising, it poses a lot of problems. Firstly, it is difficult for all the nodes in large P2P networks to trust the central party. Secondly, as the central server is involved in each and every step, it could be overloaded with too many service requests. Thirdly, the central server becomes the single point of attack and if by any means if it is compromised, the whole system would collapse and it might become impossible to detect Sybils in the network.

**(iii) Random Key pre-distribution**
A promising key distribution technique in wireless sensor networks is random key pre-distribution [5]. It assigns a random set of keys or key related information to every node in the network. During the initialization phase, if a set of keys of a node match with another, secure links can be formed. These keys can be used for secure communication between the nodes.

Distributed key pre distribution is based on the assumptions that (i) Each and every node in the network is associated with the key assigned to it. (ii)The network should be able to authenticate the set of keys that an identity claims to have. Even though, if an adversary tries to construct a bogus identity and relate it to some random identities keys, the bogus identity fails to get through the key validation phase. Therefore, random key pre-distribution technique turns out to be an effective solution to Sybil attacks.

**(iv) Sybil Infer**
Sybil Infer [6] is a centralized approach used to mitigate Sybil attacks in social networks. The process starts from a known trusted node, taken as reference, and then Sybil probability is assigned to each node using Bayesian Inference. In other words, it assigns the rank to each node which is nothing but the degree of Sybil certainty. The overall time complexity of Sybil infer is $O(V\log V)$, where V is the number of nodes or vertices in the social graph. Sybil infer is suitable for networks which holds only up to 30K nodes and it is not scalable to larger networks.

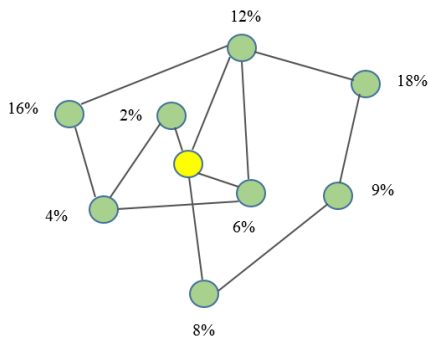

*Figure 3: Sybil probabilities with reference to a trust node*

**(v) Sybil Guard and Sybil Limit**
Sybil guard [1] is a robust protocol, which reduces the impact of Sybil attacks in social networks. Sybil guard considers the P2Por node-to-node structure in social networks to overcome these attacks. In social networks, each user is considered as a node and we assume that all the nodes (honest and Sybil) together form the social graph. In social networks, an edge is said to be formed if a relationship exists between two nodes. If a relationship is formed between an honest node and a Sybil node, the edge is called as an Attack edge [18]. Generally, the numbers of attack edges are limited as relationships are generally formed based on trust and prior communication. As the number of attack edges or relationships between honest nodes and Sybil nodes are limited, there exists a small gap or a small cut between honest region and the Sybil region. This principle is used to identify the Sybil nodes in social networks. This is because most of the nodes generally tend to remain in their own community, the honest nodes remain in the honest region and the Sybil nodes remain in the Sybil region.

In Sybil guard, we assume that we know a trusted node and trusted path [6]. From this trusted node, *m* random paths are established. We call these paths as *verifiers*. Similarly from a suspect node also, *m* random paths are established. If a path from suspect node intersects with the verifier, it is said to be *verified*. When majority of the paths emerging from a suspect node are verified, it is said to be honest node otherwise it is called as a Sybil node. Suppose, if a Sybil path crosses the small edge and meets the *path verifier* many times, it may be noted as honest node. However, as the number of attack edges are limited, i.e. the relationships between honest nodes and Sybil nodes are limited, the Sybil initialized paths will not be verified generally.

Sybil limit [2] works on the same principle of Sybil guard [1]. The only difference between the two is, Sybil guard assumes $O(\sqrt{n} \log n)$ as the lower bound for accepting Sybil nodes whereas Sybil limit improves the factor by considering $O(\log n)$ Sybil nodes as lower bound for accepting Sybil nodes, where *n* is the total number nodes or users in the social network graph (a graph of social network with nodes being the vertices) [4]. The main limitation of Sybil guard and Sybil limit is that it cannot detect more than one Sybil node at a time.

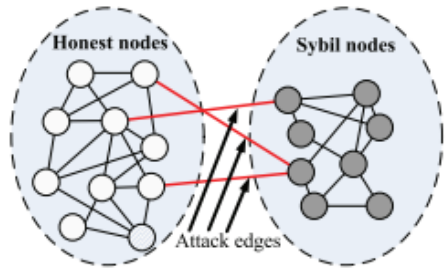

*Figure 4: Attack edges between honest region and Sybil region [6]*

**(vi) Sybil Defender**
Sybil defender [31] is one of the famous anti-Sybil detection schemes which use community detection approach. It also uses random walkers like Sybil Guard [1] and Sybil Limit [2] but uses minimum number of random walks when compared to them. The algorithm is divided in two parts (i) To find a Sybil node, (ii) To find the region surrounding the Sybil node which is assumed to be the Sybil community. The assumption is built on the fact that the numbers of attack edges are limited, i.e. the number of Sybil identities are negligible when compared to honest identities, and also the honest region is fast mixing, i.e. nodes in the honest region converge very fast when compared to Sybil region.

To detect a Sybil node we start from a known honest node and find the k-hop neighbors which are considered as honest nodes. These k-hop neighbors are considered to be honest, based on the assumption that honest region is fast mixing. These nodes serve as the boundary for the honest region. From each of these honest nodes, a large number of random walks with varying lengths are performed and the frequency distribution of nodes is observed. Frequency is defined as the number of times a random walk traverses through a node. From the suspect node also, with varying length of random walks the frequency distributions of nodes is observed. If both the distributions are same, the node is considered to be honest, otherwise it is called a Sybil node.

Once a Sybil node is detected, the Sybil community is found based on the intuition that Sybil nodes generally tend to connect with other Sybil nodes only. In Sybil community detection, partial random walks are used instead of original random walks. Partial random walks are same as normal random walks except that, a node once traversed can't be traversed more than once. Taking Sybil node as reference, a partial random walk is initiated, which reaches a node where all of its surrounding nodes are traversed and the path is dead. After performing a large number of partial random walks, it yields many dead paths. Traversing through these dead paths yields the Sybil community.

**(vii) Symon**
Symon [46] is a novel approaches to defend against Sybil attacks in distributed decentralized networks. It ensures honest nodes are protected from Sybils with high probability. In Symon, each and every node in the network is associated with a non-Sybil called as Symon. The non-Sybil or Symon is chosen dynamically, such that the chance of both nodes being Sybils is impossible. Each Symon is given the responsibility of monitoring the activities of the given node making it impossible for the other node to compromise the system. A Symon should make sure that a malicious node has to invest a lot of cost to create a bogus identity. In addition, in Symon approach any node in the network can verify with high probability whether a pair nodes are Sybils or not.

**(viii) Sybil Shield**
Social network based anti-Sybil schemes came into prominence due to the limitations posed by central trust authority and resource testing schemes. It uses the trust relationships in social networks to minimize the influences of Sybil attack. Sybil guard [1] is the first approach which used social network architecture to reduce the Sybil influences.

Unlike previous social network based approaches, which assumes that the social network graph is divided in two regions namely honest and Sybil regions, Sybil shield [46][47] assumes that there exists multiple number of large, medium and small communities in real-time social networks. In [47], a real-time experiment conducted on Myspace reveals that it can be divided into 19 communities, with the smallest size being 12 and the largest size of 33,887 nodes with a number of foreign edges between communities. An edge between two different communities is called a foreign edge. It also assumes that social networks are fast-mixing i.e. even though if an adversary creates large number of Sybil identities, the number of trust relationships established between an honest node and a Sybil node is limited. Therefore, foreign edges formed between honest communities are more when compared to honest and Sybil communities.

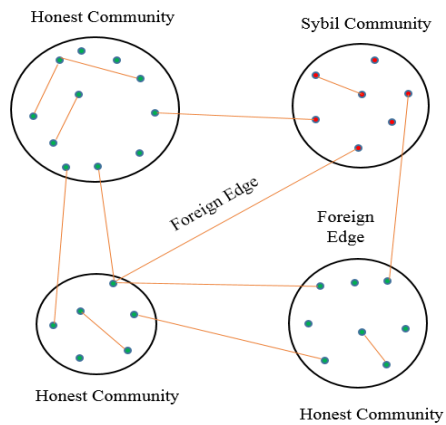

*Figure 5: Different communities in real world Social Networks*

In Sybil guard, the chances of an honest suspect node being treated as Sybil is more, as the approach fails to distinguish between honest and Sybil communities. Sybil shield aims to increase the acceptance rate of an honest suspect node by using the random route approach. Random routes are nothing but a modification of random walks. In Sybil shield, a suspect node is considered to be honest, if the random routes of verifier and suspect intersect. Unlike, other schemes which rejects a node, if their random walks doesn't intersect, Sybil shield gives a second chance for the suspect nodes using agent walk approach. Agents are nodes chosen from other communities, other than the verifier's community, to verify whether a suspect node is Sybil or not. Agents are selected carefully along the edges of the verifier and based on the assumption that a random route initiated from a verifier has more probability to land in honest community than Sybil community. If at least a certain number of random routes from agents intersect with suspect node, then the node is considered as honest otherwise it is considered as a Sybil node.

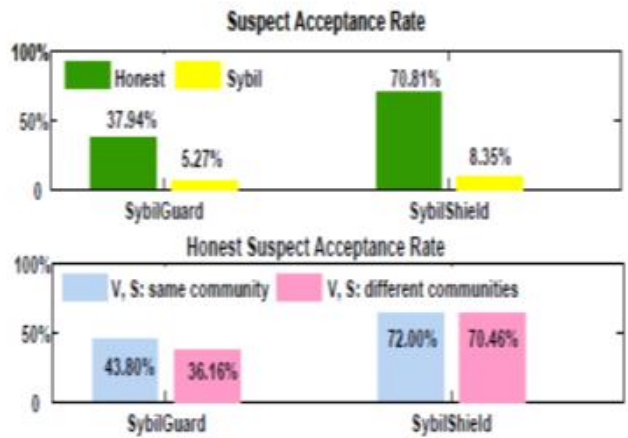

*Figure 6: Honest suspect nodes acceptance rate* [46]

**Behavior based Sybil Detection**
Most social network based Sybil detection schemes work on the random walk approach and assume that most of the Sybil identities generally establish connections only with Sybil identities and tend to remain in their own community. A recent study in [16][48] online social network Renren, states that the number of social connections established between Sybil users and honest users is growing exponentially and an experiment [49] states that in a social network Sybil users generally won't establish connections with other Sybil users. In real time social networks, considering the pace at which Sybil identities are multiplying, social network based approaches alone wouldn't be sufficient to detect the Sybil attacks which led to many novel anti-Sybil approaches based on user's behavior.

In behavior based anti-Sybil schemes, the users browsing and clicking habits are considered to separate Sybil users from normal users [48].The general activities performed by a user in online social networks are uploading pictures, viewing other's photos and profiles, sending messages, sharing content, playing games, posting blogs and commenting on the blogs. Unlike normal users, Sybil users activities are quite narrow such as befriending (Send friend requests and establish connections), sending messages and sharing content with others which are considered to be the main sources to spread spam and malware in social networks.

Considering the click transitions as a Markov chain [51], a random process which changes from one state to another, and assuming each click pattern as a different state, it can be observed that normal user's behavior is unpredictable and changes from one state to another randomly whereas Sybil users participate only in specific activities repeatedly. The Support Vector Machine approach (SVM) has been proposed [50],[53] to monitor specific features such as clicks in particular time period, number of clicks in a session, time interval between two clicks and other features which yielded high results to detect Sybil activity. Another study proposes [48] three models namely click sequence model, time-based model and hybrid model, which combines similar click patterns for analyzing the users behavior. Based on some standard similarity measurements, a graph is constructed which is used to detect Sybil users. In [52], Wang et al, proposes a Sybil detection scheme based on crowd-sourcing and Social Turing test. Turing test is the ability of a machine to think intelligently similar to a human being. Even though, if an attacker has a wide range of tactics, she cannot pass the turing test which increases the detection efficiency. In addition, crowdsourcing provides a platform for users to share their opinions, ideas and provide ratings to improve the Sybil detection accuracy.

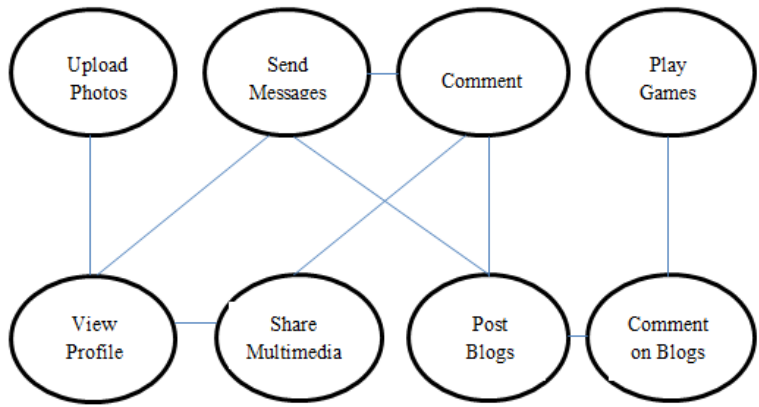

*Figure 7: State transitions for Normal Users*

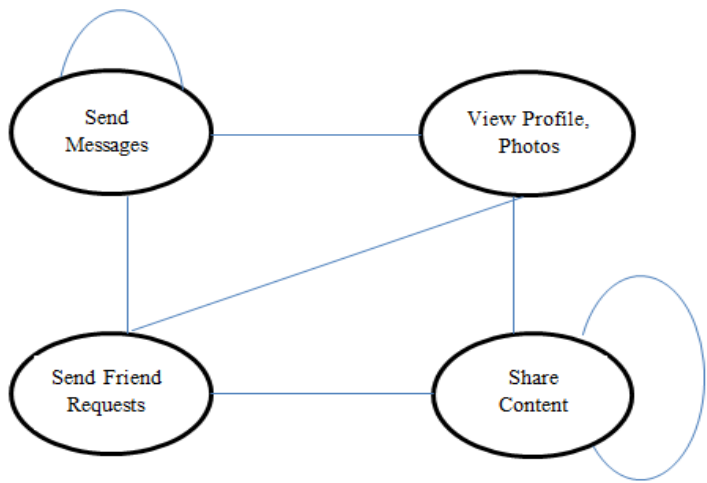

*Figure 8: State transitions for Sybil Users*

**Conclusion**
Sybil attack is most crucial in distributed decentralized systems like social networks. Since too much personal and sensitive information is posted openly in social networks, the threats posed by Sybil attacks are very serious. In this paper, we discussed the need for social network security and the effects of Sybil attacks in social networks, classification of Sybil attacks to understand about the sources of the attack, examples of systems vulnerable to Sybil attacks and a summary of the current trends in Sybil defenses. The Sybil defenses are classified according to the attacker's capability and behavior attributes of Sybil user's and advantages and disadvantages of each method is discussed.